\documentclass[%
 reprint,
 amsmath,amssymb,
 aps,
 floatfix,
]{revtex4-1}

\usepackage{graphicx}
\usepackage{dcolumn}
\usepackage{bm}

\begin{document}

\preprint{APS/123-QED}

\title{Review of Pulsar Timing Array for Gravitational Wave Research}

\author{Pravin Kumar Dahal}

\affiliation{
 Dept. of Physics and Astronomy, Macquarie University, North Ryde, NSW, 2109, Australia
}

\date{\today}

\begin{abstract}
 Ongoing research on Pulsar Timing Array (PTA) to detect gravitational radiation is reviewed. Here, we discuss the use of millisecond pulsars as a gravitational wave detector, the sources of gravitational radiation detectable by PTAs and the current status of PTA experiments  pointing out the future possibilities.
\end{abstract}

\maketitle

\section{Introduction}
Radio pulsars, simply known as pulsars are highly magnetized rapidly rotating neutron stars with a coherent source of radio waves. Milliseconds pulsars are the special class of pulsars with a stable rotational period of about 1-10 milliseconds and thus significantly stable pulse frequency.  Because of their high timing accuracy, observations of a group of pulsars started known as pulsar timing array (PTA) program. The PTA program has developed into Parkes Pulsar Timing Arrays and is currently taking observations of 25 pulsars \cite{2}. Along with this, the North American Nanohertz Observatory for Gravitational Waves (NANOGrav) is taking observations of 45 pulsars\cite{6} and 42 pulsars are being observed by European Pulsar Timing Arrays (EPTA)\cite{40}. The collaboration of all these three PTAs is called International Pulsar Timing Array (IPTA)\cite{16} and their new data release consists of 65 pulsars\cite{40}. The first discovery of gravitational wave by LIGO/ VIRGO collaboration from binary black hole mergers opened the field of gravitational wave astronomy\cite{17}. Owing to the short arm length of earth-based detectors like LIGO, they are able to detect the gravitational wave of high frequency in the range from $10-10^4$ Hertz. Pulsar timing arrays are able to detect the gravitational wave of low frequency in the range from $10^{-9}-10^{-7}$ Hertz. This is because the Earth-pulsar system acting as a detector in PTA has a huge arm length. There are space-based detectors like Laser Interferometer Space Antenna (LISA), that are able to detect the gravitational wave of medium frequency, in the range between LIGO and PTAs (i.e. $10^{-4}-10^{-1}$ Hertz)\cite{24}. While Earth-based detectors detect a burst of waves from stellar-mass objects just before merging, PTAs detect waves from supermassive black holes in the early stage of inspirals. So PTAs provide a view of the gravitational wave sky complementary to the earth-based and space-based detectors. This makes them useful to uncover the mysteries of galaxies formation and black hole dynamics\cite{25}. Similarly, we could obtain a better estimate of the galaxy merger rate and the population of supermassive black hole binaries in the Universe\cite{26}. PTAs also provide an opportunity to test the theory of gravitation in nanohertz regime\cite{23}

Besides gravitational waves detection, other applications of PTAs include providing time standard for long time scales and measurement of solar system ephemerides\cite{18} and a better understanding of the properties of the interstellar medium\cite{45}. The theory of detection of the gravitational wave using PTA is discussed in Sec.\ref{s2}, possible sources of gravitational wave detectable by PTAs are discussed in Sec.\ref{s3} and Sec.\ref{s4} is the concluding remark with the present status of PTAs and their future targets. Readers are directed to the reviews by Lommen\cite{42}, Tiburzi\cite{43}, Burke-Spolaor et.al.\cite{44} and the references therein for comprehensive discussions on the topic.

\section{Detection of Gravitational Wave Using PTA}\label{s2}
When a gravitational wave passes between the Earth and pulsar system, the time of arrival of the pulsar signal from the pulsars changes. This induced frequency change due to the gravitational wave is given by \cite{1}:
\begin{equation}
    \delta \nu/\nu=-H{^{i j}}[h_{i j}(t_{ e},x^i_{ e})- h_{i j}(t_{p},x^i_{p})]
    \label{1}
\end{equation}
where $H^{i j}$ depends on the angle between the earth, pulsar and the source of gravitational wave. $h_{i j}$ is the dimensionless amplitude of gravitational wave at the earth (at position $\vec{x_e}$ and time $t_e$) and at the pulsar (at position $\vec{x_p}$ and time $t_p = t_e-D/c$, $D$ being the distance between the earth and the pulsar) as shown in Fig.\ref{fig1} (note: the origin of our coordinate system is the solar system barycentre):
\begin{figure}[bth]
\begin{center}
\includegraphics[scale=0.9]{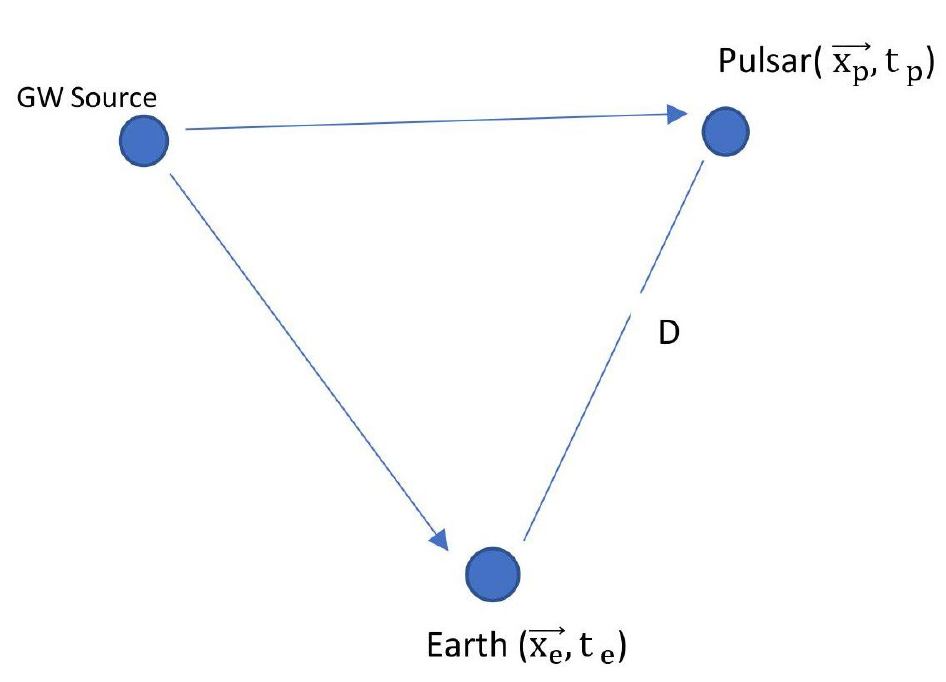}
\caption{system of Earth, Pulsar and GW}
\label{fig1}
\end{center}
\end{figure}
This variation in pulse frequency due to gravitational wave appears as an anomalous residual in pulse arrival time, and is given by:
\begin{equation}
    R(t)=-\int_{0}^{t}\frac{\delta\nu}{\nu}dt
\end{equation}
 PTA involves analysis of the set of pulsars to look for the correlation in arrival times of pulses emitted by them. This correlation is contributed by the gravitational wave strain $h_{i j}(\vec{x_e}, t_e)$ at the earth and not by $h_{i j}(\vec{x_p}, t_p)$ at the pulsar. In timing single pulsar, a stochastic signal is picked up as a timing noise, which necessitates the timing of an array of pulsars to segregate the noise from these signals.
 
 For correlation analysis of pulse signal from pulsars, let us write Eq.\ref{1} as:
 \begin{equation}
     \frac{\delta\nu_i}{\nu}=\alpha_i h(t)+n_i(t)
 \end{equation}
where $h(t)$ is gravitational wave strain and is common to all pulasars, $\alpha_i$ is geometric term depending the orientation of pulsars and $n_i(t)$ represents the noise of particular pulsar. Cross-correlation of this frequency variation from two pulsars gives:
\begin{eqnarray}
    c_{i j}(\tau)=\alpha_i\alpha_j<h^2>+\alpha_i<h n_j>+\alpha_j<n_i h>\nonumber\\
    +<n_i n_j>
    \label{3}
\end{eqnarray}
where, $<h^2>=\frac{1}{T}\int_{-T-\tau}^{T+\tau}h(t)h(t+\tau)dt$; is the time average of $h^2$. Here, $T$ is total data span time and $\tau$ is the time lag in receiving the signal from second pulsar. If the distribution of gravitational radiation is assumed to be isotropic, then $<h^2>$ is independent of direction and we have the average of angular factors $\alpha_i$ $\alpha_j$ as:
\begin{equation}
    \alpha_{i j}=\frac{1}{4\pi}\int\alpha_i\alpha_j d\Omega
\end{equation}
In the limit of large data span time $T$, it is assumed that the noise from two pulsars $n_i$ and $n_j$ and the gravitational wave strain $h(t)$ all are uncorrelated to each other causing all the terms, except first, on the right hand side of Eq.\ref{3} to vanish. So, we have:
\begin{equation}
    c_{i j}(\tau)=\alpha_{i j}<h^2>+\delta c_{i j}
    \label{4}
\end{equation}
where $\delta c_{i j}$ is an estimation error in infinite $T$ and,
\begin{equation}
    \alpha_{i j}=\frac{1-cos\theta_{ij}}{2}ln\Big(\frac{1-cos\theta_{ij}}{2}\Big)-\frac{1}{6}\frac{1-cos\theta_{ij}}{2}+\frac{1}{3}
\end{equation}
$\theta$ being the angle between two pulsars. The plot of this correlation function with angle is commonly known as Hellings and Downs curve and is shown below in Fig.\ref{fig2}.
\begin{figure}[bth]
\begin{center}
\includegraphics[scale=0.35]{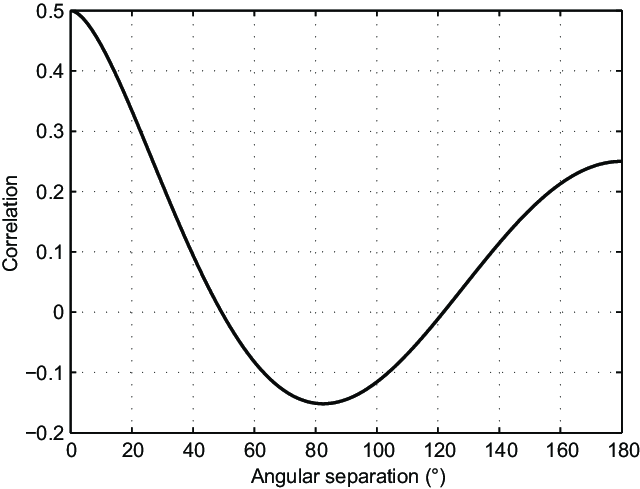}
\caption{The Hellings-Downs curve}
\label{fig2}
\end{center}
\end{figure}
From figure, we can see that the unique deformation produced by the gravitational wave is solely dependent on the angular separation $\theta$. The correlated signal in the data from continued timing of at least three pulsars non-coplanar to the solar system characterizes a gravitational wave source. Hellings and Downs used this correlation analysis from four pulsars to limit the energy density of stochastic background at frequencies below $10^{-8}Hz$ to be $10^{-4}$ times the critical density of the universe ($\rho_c=\frac{3H_0^2}{8\pi G}$) \cite{3}. The discovery of millisecond pulsars improved this result of energy density of background up to $10^{-6}$ times the critical density \cite{4}. 

\begin{figure}[bth]
\begin{center}
\includegraphics[scale=1.10]{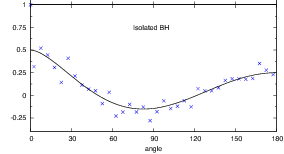}
\caption{Plot of Correlation function with an angle between pulsar pair for a black hole binary. This figure, drawn assuming 100 randomly distributed pulsars, is taken from Ref.\cite{21}}
\label{fig3}
\end{center}
\end{figure}

As seen above, the standard tensor correlation method for PTA data analysis assumes isotropic distribution of gravitational wave signal with Gaussian distribution. This assumption is justified if the number of sources (black hole binaries) emitting particular frequency bins are large enough to make signal distribution Gaussian. However, recent models on the black hole population show that the gravitational wave signal from black holes binaries could be anisotropic and dominated by some nearby sources\cite{20}. So, our analysis for isotropic signal distribution should not necessarily hold if this is the case. Cornish and Sesana\cite{21} considered the gravitational wave signal from the single black hole binary and showed that the correlation relation from Hellings-Downs (Eq.\ref{4}) continues to hold for anisotropic signal distribution given the number of pulsars are sufficiently large. The reason behind this is attributed to the quadrupolar nature of the gravitational wave. Fig.\ref{fig3} shows the correlation curve for isolated black holes binary depicting the nature of the curve following the Hellings-Downs curve of Fig.\ref{fig2}.
Cornish and Sampson\cite{19} discussed on the reduction in detection probability of non-Gaussian/ anisotropic signal by timing the limited number of pulsars. To detect a gravitational wave from PTAs, two conditions must be satisfied: firstly, the amplitude of gravitational wave should be large enough such that it is statistically significant and secondly, the gravitational wave frequency should lie within the frequency range sensitive for PTAs.

\section{Gravitational Wave Sources Detectable by PTAs}\label{s3}
The nature of gravitational waves depends on the sources producing them and this determines the graph of timing residual defined above. Some of the possible candidates of gravitational waves that are detectable by PTAs are:
\subsection{Stochastic Backgrounds}\label{A}
The stochastic background is due to the incoherent superposition of randomly emitted waves from a large number of weak, unresolved and independent sources. It includes the gravitational waves from a wide range of cosmological phenomena including cosmological phase transition\cite{29}, cusps, and kinks of cosmic strings\cite{27} and cosmic inflation\cite{28} and the waves from astrophysical phenomena like the coalescence of massive black holes\cite{30}. Although all of these background sources fall in the PTA range, the background signal from supermassive black holes binaries is expected to dominate in amplitude.  So, we constrain our discussion on the spectrum of coalescing black holes and the characteristic strain (i.e. gravitational wave amplitude) of this phenomena is given by:
\begin{equation} 
   h_c(f)=A f^{-\frac{2}{3}}
\end{equation}
where, A is the dimensionless amplitude at reference frequency $f=1 yr^{-1}$ and its predicted value is $10^{-15}$ \cite{9}. This value defines the dynamical model for the evolution of supermassive black holes binaries and is being constrained strictly by the recent PTA observations\cite{6, 22}

We will now discuss a frequentist formalism developed by Rosado et al.\cite{10} to compute detection probability of stochastic background as a function of observation time. In general, detection is a probabilistic endeavour since we have to rely on statistics for the realization of true faint signal over noise. In PTA experiment, we measure the strength of signal (proportional to $A^2$ where $A$ is the gravitational wave amplitude discussed above) as cross-correlation between two pulsars. Although we do not have access to $A_{true}$, we can experimentally measure upper limit $A_{ul}$. Hence, using Bayes theorem, we can obtain $P(\frac{A_{true}}{A_{ul}})\propto P(\frac{A_{ul}}{A_{true}})P(A_{true})$. $A_{true}$ can be estimated from the theory and  $P(\frac{A_{ul}}{A_{true}})$ can be calculated, as will be shown below, to determine $P(\frac{A_{true}}{A_{ul}})$ (ref.\cite{12} shows the plot of $P(\frac{A_{true}}{A_{ul}})$ with $A_{true}$ for $A_{ul}$ given by PPTA and NANOGrav).

Unlike the tensor correlation approach, the frequentist scheme assumes a single number, say X, that contains both the noise and the gravitational wave signal. In this scheme, a higher amplitude $A$ yields higher X with fluctuations from all possible noises. Hence, setting an upper limit $A_{ul}^{95\%}$ in this scheme would imply $A_{true}^{95\%}$ with a $95\%$ realization of noises. To determine the upper limit of gravitational wave background signal from pulsars widely separated in the sky, cross-correlation statistics, as discussed above is implied.

Firstly, assume that, in the absence of gravitational wave background, the cross-correlation of all noise processes (i.e. the strength of noise signal) follows Gaussian distribution with mean $A_{B}^2$ and standard deviation $\sigma_B$:
\begin{equation}
    P(A^2/A_B)={\frac{1}{\sqrt{2\pi\sigma_B^2}}exp\Big[\frac{-(A^2-A_B^2)^2}{2\sigma_B^2}\Big]}
\end{equation}
We similarly assume that the gravitational wave signal present in the data follows Gaussian distribution with different mean $A_{true}^2$ and standard deviation $\sigma_0$:
\begin{equation}
    P(A^2/A_{true})={\frac{1}{\sqrt{2\pi\sigma_0^2}}exp\Big[\frac{-(A^2-A_{true}^2)^2}{2\sigma_0^2}\Big]}
\end{equation}
Given the threshold amplitude $A_{ul}$, the intgral of background signal overall values of $A_{ul}$ gives false alarm probability ($\alpha$):
\begin{equation}
 \alpha = \int_{A_{ul}^2}^{\infty}P(A^2/A_B)dA^2=\frac{1}{2}erfc\Big[\frac{A_{ul}^2}{\sqrt{2}\sigma_B}\Big],
\end{equation}
assuming the noise has zero mean i.e. $A_B=0$. The integral of true signal overall values of $A>A_{ul}$ gives detection probability ($\gamma$):
\begin{equation}
    \gamma = \int_{A_{ul}^2}^{\infty}P(A^2/A_{true})dA^2=\frac{1}{2}erfc\Big[\frac{A_{ul}^2-A_{true}^2}{\sqrt{2}\sigma_0}\Big]
    \label{5}
\end{equation}
where, erfc is error function.
False alarm probability of $\alpha_0=0.1\%$ corresponds to $3\sigma$ detection and in that case, detection probability is:
\begin{equation}
     \gamma=\frac{1}{2}erfc\Big[\frac{\sqrt{2}\sigma_{B}erfc^{-1}[2\alpha_0]-A_{true}^2}{\sqrt{2}\sigma_0}\Big]
     \label{6}
\end{equation}
The relation for cross-correlation i.e. the measured strength of the gravitational wave signal is given by Eq.\ref{4}. In contrast to the cross-correlation given in Ref.\cite{10}, the filter function is missing in our relation. The reason behind is that Eq.\ref{4} is derived under the assumption of non-deterministic and isotropic signal. As discussed in sec.\ref{s2}, this assumptions is valid in two conditions: first, when the number of sources emitting gravitational radiation are independent and infinite and second, when the number of pulsars are infinite. We also know that the detection probability is maximum when both the cases are satisfied. So, the Eq.\ref{4} for cross-correlation is the one such that the detection probability is maximum (any forms of noise are ignored). Using the relation for cross-correlation, $\sigma_0$, $\sigma_1$, and $A_{true}$ can be determined (see Ref.\cite{10}) to calculate the detection probability given in Eq.\ref{6}. Plots of detection probabilities with background wave amplitude ($A_{true})$ and observation time for various PTAs are shown in Ref.\cite{12}.

Now, from Eq.\ref{5}, we have:
\begin{equation}
    A_{ul}^2=A_{true}^2+\sqrt{2}\sigma_0 erfc^{-1}(2\gamma)
\end{equation}
From this equation, we can infer that, $P(\frac{A_{ul}^2}{A_{true}})$ follows Gaussian distribution with mean $A_{ul}^2=A_{true}^2+\sqrt{2}\sigma_0 erfc^{-1}(2\gamma)$ and variance $\sigma_0^2$. From this, we can find $P(\frac{A_{ul}}{A_{true}})=2A_{ul}P(\frac{A_{ul}^2}{A_{true}})$ enabling us to evaluate $P(\frac{A_{true}}{A_{ul}})$.

To estimate the gravitational background from black hole mergers, one should understand the mechanism for the formation of binaries to know the overall merger rate and the merger rate as a function of redshift. Jenet et al.\cite{5} calculated the detection significance using the correlation method and came up to the conclusion that the probability of detecting stochastic gravitational wave is approximately $95\%$ by using $40$ pulsars with timing precision of $100 ns$, observed $250$ times for over $5$ years. Using Eq.\ref{4}, Rosado et al.\cite{10} have computed the detection probability of background signal from IPTA for the first 10 years to be approximately $37\%$. The gravitational wave background is undetected until now and PTAs are starting to constrain the limits on the background signal. Recently released $11-$year dataset from NANOGrav\cite{6} claims to have placed a $95\%$ upper limit on gravitational wave amplitude $A<1.45\times10^{-15}$  and this is starting to question at least one of the assumptions underlying our model on the formation of gravitationally bound supermassive black holes binaries\cite{2}. These results can be combined to conclude that the PTA's consisting of few pulsars could provide the stringent upper limit, but is insufficient to give satisfactory result for the detection probability. Cornish and Sampson\cite{19} showed a reduction in detection probability because of having a finite number of pulsars and limited gravitational wave sources using correlation analysis. The reason is attributed to the breaking of statistical isotropy of gravitational wave signal assumed in the derivation of Eq.\ref{4}

\subsection{Continuous Waves From Individual Binaries}\label{3b}
Another strongly anticipated source of gravitational waves in PTA frequency range is individual nearby sources of Super Massive Black Hole (SMBH) binaries which emits sufficiently strong continuous gravitational waves. For SMBH binary, assumption of low eccentric orbit and the evolution solely by energy loss via gravitational radiation leads to the equation for characteristic strain amplitude \cite{7}:
\begin{equation}
h_c={\Big(\frac{128{\pi}^{1/3}}{15}\Big)}^{1/2}\frac{{\mathcal{M}}^{5/3}}{r}f^{2/3}
\end{equation}
where, $\mathcal{M}=\frac{(m_1m_2)^{3/5}}{(m_1+m_2)^{1/5}}$ is the chirp mass (effective mass of the binary that determines the strength of gravitational wave emitted), $r$ is the luminosity distance and $f$ is the frequency of the gravitational wave. For a given $M$, the chirp mass $\mathcal{M}$ will be maximum when $m_1=m_2$ i.e. when the masses of two black holes of the binary are equal and in that case:
\begin{equation}
    h_{c,max}=1.54\frac{M}{r}f^{2/3}
\end{equation}
where $M=m_1+m_2$ is the total mass of the binary. $h_{c,max}$ being the maximum value of the amplitude of gravitational wave emitted by SMBH binary of total mass $M$, PTA must be sensitive to this strain before beginning the hunt for continuous waves in particular galaxies.

To calculate the detection probability of these waves, Ellis et al.\cite{11}have presented a derivation of $\mathcal{F}$-statistics for waves from individual sources. $\mathcal{F}$-statistic is the likelihood function maximized with respect to the parameters of the signal. It was first developed by Jaranowski et al.\cite{31} for the search of gravitational wave signal from spinning neutron star for LIGO. In this statistic, if the maximum of the likelihood function is greater than some threshold determined by noise, detection is said to be made. In the absence of gravitational waves, $\mathcal{F}$-statistics is a $\chi^2$ distribution with probability distribution function:
\begin{equation}
    P_0(\mathcal{F})=\frac{\mathcal{F}^{\frac{n}{2}-1}}{(\frac{n}{2}-1)!}exp(-\mathcal{F})
\end{equation}
where, $n$ is the degrees of freedom of the distribution. Similarly, if gravitational waves is present, then the statistics is non-central $\chi^2$ distribution with probability distribution function:
\begin{equation}
    P_1(\mathcal{F, \rho})=\bigg(\frac{2\mathcal{F}}{\rho^2}\bigg)^{\frac{n}{4}-\frac{1}{2}}J_{\frac{n}{2}-1}(\rho \sqrt{2\mathcal{F}})exp(-\mathcal{F}-\rho^2/2)
\end{equation}
where, $J$ is modified Bessel's function of first kind and non-centrality parameter $\rho$ is equal to optimal signal to noise ratio. We have, for $N$ pulsars, degrees of freedom $n=2N$\cite{41}. Now, assuming that we know the intrinsic parameters of the signals we are searching for, we can calculate false alarm probability by integrating probability density function in the absence of signal as:
\begin{equation*}
    \alpha_i=\int_{\overline{\mathcal{F}}}^{\infty}P_0(\mathcal{F})d{\mathcal{F}}
\end{equation*}
where, $\overline{\mathcal{F}}$ is the threshold of detection. If the intrinsic parameters are not known, then the false alarm probability $\alpha$ is:
\begin{equation}
    \alpha=1-(1-{\alpha}_i)^{N_c}
\end{equation}
$N_c$ being the number of independent cells in parameter space. We have seen from Eq.\ref{1} that the timing residual does depend on the gravitational strain amplitude at the earth and the pulsar. Both the Earth term and pulsar term of the signal from individual sources are known to follow ${\mathcal{F}}$-statistic\cite{11}. But pulsar term is negligible either when the number of pulsars is large or when all of the pulsars terms are at different frequency bins than at the Earth term. In that situation, the number of independent cells $N_c$ can be approximated to the number of templates used in the search to determine $\alpha$\cite{39}.

As discussed in Sec.\ref{A}, it is customary to fix $\alpha={\alpha}_0$ to obtain the threshold $\overline{\mathcal{F}}$, which allow us to calculate the detection probability ${\gamma}_i$ from numerical integration:
\begin{equation}
{\gamma}_i=\int_{\overline{\mathcal{F}}}^{\infty}P_1(\mathcal{F}, \rho)d{\mathcal{F}}
\end{equation}
This is the probability of detecting binaries in particular frequency bin. The total probability of detecting at least one binary in all frequency bins is:
\begin{equation}\label{8}
    \gamma=1-\prod_i(1-\gamma_i)
\end{equation}
where, index $i$ include all frequency bins in the range. Eq.\ref{8} gives the detection probability of continuous waves from individual sources given the value of signal to noise ratio $\rho$ (see Ref.\cite{10} for the calculation of $\rho$).

The 11-years data of NANOGrav from the sample of $45$ pulsars placed an upper limit on the gravitational strain of $h_c<7.3\times10^{-15}$ at $95\%$ confidence level\cite{32}. From the upper limit, they have placed constraints on the population of supermassive black holes binaries with particular chirp mass. Similarly, PPTA placed an upper limit of $h_c<1.7\times10^{-14}$ from $20$ pulsars observation and EPTA reported the limit of $h_c<1.3\times10^{-14}$ from the observation of 42 pulsars both at $10nHz$. These upper limits on $h_c$ from PPTA and EPTA have been analyzed in \cite{8} to constrain the mass ratios of black hole binaries in galaxy samples. The detection probability of continuous waves from individual sources is calculated by Rosado et al.\cite{10} to be about 10-20\% after approximately 15 years from now. These figures again allow us to conclude that PTA's consisting of few pulsars are sufficient to place the stringent upper limit, but are insufficient to provide satisfactory detection probability.

\subsection{Gravitational Waves From Burst Events}
Bursts events produce transient signals and being sensitive to the initial conditions their nature can vary widely. Some of the burst events detectable by PTAs are the formation of supermassive black holes, black holes binaries rotating in highly eccentric orbits and encounters of massive objects. Final stage of inspiral of supermassive black holes mergers\cite{33}, asymmetric supernovae\cite{35} and encounter of massive objects\cite{36} can cause permanent distortion in spacetime called `Memory Events'. Some of these burst events are within the current sensitivity range of PTAs. So, we will now focus our discussion on the detection of these events.

We have discussed in Sec.\ref{3b} that the gravitational wave amplitude from an individual binary is maximum when the mass of the constituting black holes are comparable. So, the gravitational wave amplitude of `+' polarized wave from the black hole binary contributing to the memory event is given by\cite{54}:
\begin{equation}
    h^{mem}_+=\frac{1}{24 r} \sin^2\Theta (17+\cos^2\Theta)\Delta E_{rad}
\end{equation}
where $r$ is the luminosity distance, $\Theta$ is the inclination angle of the binary just before merger and
\begin{equation}
    \Delta E_{rad} \simeq \bigg(1-\frac{\sqrt{8}}{3}\bigg)\frac{G \mu}{c^2}
    \label{22}
\end{equation}
is the energy radiated during merging in leading order approximation\cite{55}. Here, $\mu$ is the reduced mass. The contribution to the $h^{mem}_+$ is maximum when the black holes in the binary have comparable masses and that is when burst event is most likely detectable by PTAs. Gravitational wave amplitude for cross polarization $h^{mem}_\times$ vanishes for circular binary. For a black hole merger, each with $10^9 M_{\odot}$ mass at a distance of $1Gpc$, the expected gravitational wave amplitude $h^{mem}_+$ from Eq.\ref{22} is approximately $10^{-15}$\cite{34}.

The memory events are undetected till now, but PTAs could place the upper limit on these events to provide useful information regarding supermassive black holes binaries population\cite{34}. To see this, we assume that the burst signal follows a Poisson distribution with probability density function:
\begin{equation}
    P(h)=1-e^{-\Lambda(h) t}
\end{equation}
where $\Lambda(h)$ is the rate of signal detection for some amplitude $h$ greater than the threshold amplitude and $t$ is observation time. $\Lambda(h)$ being the rate of burst wave detection does depend on the population of supermassive black hole binaries (this can be seen in the work of Cordes and Jenet\cite{56}). For a given merger rate, if an observation is made to some characteristic time $T$, the signal from burst event should be detected. The signal undetected until time $T$ implies that we might have overestimated the supermassive black holes merger rate. The constraint on the merger rate can be given as:
\begin{equation}
    \Lambda<-\frac{1-P}{T}
\end{equation}
assuming $\Lambda$ is constant for all amplitude $h$ above the threshold. NANOGrav, during its first five years, has placed an upper limit on the rate of burst with memory events of different amplitudes\cite{38}.

\section{Results and Conclusion}\label{s4}
\begin{figure}[bth]
\includegraphics[scale=0.50]{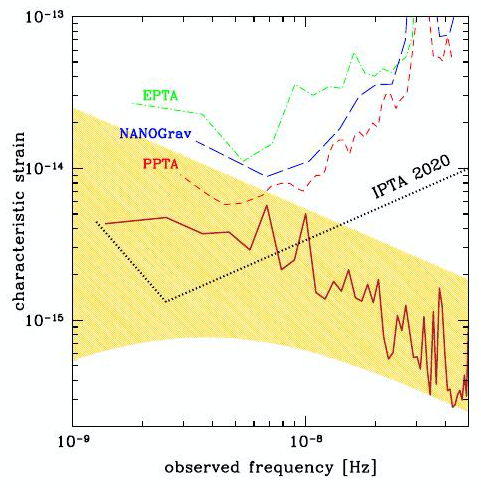}
\caption{Plot of bounds on strain sensitivity from different PTAs (dotted coloured lines) with frequency. The dotted black line represent the bound IPTA will reach by 2020. The shaded region and the solid line represents the theoretical bounds for stochastic background signal from supermassive black hole binaries\cite{37}. This figure is adapted from Ref.\cite{2}}
\label{fig4}
\end{figure}

Upper limits on strain amplitude of gravitational wave signal from various PTAs along with the theoretical predictions are shown in Fig.\ref{fig4}.
PTA's constraint on upper limit of strain amplitude ($A \sim 10^{-15}$) on background signal suggests that either we might have overestimated the binary merger rate or our understanding on the evolution of supermassive black holes binaries needs revision\cite{2}. It is shown in \cite{12} that it should take another $10$ years for PTA's to reach this strain sensitivity of $\sim 10^{-15}$. Moreover, they concluded that NANOGrav+, EPTA+ and IPTA+, which actually adds $4$ millisecond pulsars per year on regular NANOGrav, EPTA and IPTA, will begin to give convincing detection probability only after $5$ years of observation beyond current dataset. It is also mentioned in Ref.\cite{13} that $5$ years after the detection of background waves, individual sources are expected to be detected. Supermassive black holes of approximately $10^8M_{\odot}$ currently inspiralling in the PTA band are supposed to be the source for LISA, proposed to launch on early 2030\cite{13}. So, the prevailing uncertainties in the detection of gravitational waves from PTA's could have big implication on LISA.

Verbiest et al.\cite{48} presented the first IPTA data release in 2016 using $49$ millisecond pulsars to place the upper limit of $.7 \times 10^{-15}$ on the gravitational wave background. This more constraining value from background amplitude led Verbiest et al. to conclude that the sensitivity of IPTA is at least twice the sensitivity of individual PTAs. Thus, new collaborations that will be established with the development of PTA experimentation in South Africa, China and India are expected to have a huge impact on PTA's sensitivity. Chinese PTA will be using two major telescopes, Five Hundred Metre Aperture Spherical Telescope (FAST) and Qi Tai Radio Telescope (QTT). They, in combination, will be sensitive to the gravitational strain of $2 \times 10^{-16}$ in a few years for background signal\cite{49}. South African PTA will use the MeerKAT telescope, which is currently being used as one of the pathfinders for Square Kilometer Array (SKA)\cite{50}. Similarly, Indian PTA is using Ooty Radio Telescope (ORT) and Gaint Metrewave Radio Telescope (GMRT)for the observation of millisecond pulsars\cite{43}.

SKA would be the world's largest telescope that has the potential of finding out all the pulsars in our galaxy with the beam pointing towards us\cite{51}. It is expected to be fully started from $2025$ and immediately after that we hope to find $205$ millisecond pulsars suitable for timing\cite{52}. Because of the large collecting area of about $1km^2$ of the SKA, it will have the timing accuracy of $10ns$ if the timing error goes as the inverse square of the collecting area. Using the $100$ millisecond pulsars, each with the accuracy of $100ns$, Ravi et al.\cite{53} have calculated the detection probability of $50\%$ for continuous waves from individual sources. Now, increasing the number of pulsars to $250$ by maintaining the timing accuracy of $10ns$ increases the signal to noise ratio by approximately $16$ times. This is obtained by using the scaling relation of signal to noise ratio given in Ref.\cite{41}. SKA is thus expected to be powerful than any of its counterparts for pulsar timing.

Although millisecond pulsars have stable pulse frequency over long time, their intrinsic frequency is subjected to 'red noise' which is not completely understood. In addition, propagation through the interstellar medium could affect the pulse frequency to contribute an additional noise. This is because of the distortion of the pulse signal by the small scale variation of constituents of the interstellar medium. This effect can be minimized by using multiple telescopes for observation of different frequency bins\cite{46}. The timing residual in pulse frequency also depends on the position of the earth with respect to the solar system barycentre. Thus the noise in timing residual could arise because of the errors in solar system ephemeris. It has been reported in Ref.\cite{6} that this noise mimics as a false background signal in high precision dataset taken for a sufficiently long time. Another factor affecting the timing residual is solar wind. The timing residual induced by solar wind depends on the line of sight of observation and on observation time (for example variation of timing residual on a daily basis). This effect can be accounted to some extent by using a wide bandwidth receiver and multiple telescopes for observation\cite{47}.

By using the knowledge of gravitational waveform in nanohertz regime, we can construct a template for the expected signal, to deal with these noises. For constructing the template, we usually assume isolated source in perfect vacuum and luckily, many effects we ignore under these assumptions are small and can be neglected\cite{14}. The most anticipated signal for PTA's is gravitational wave background and as pointed out by Hellings and Downs these signals, rather than other noise sources, would cause timing residuals from pulsars at different locations to display quadrupolar pattern (correlation between timing residuals in different directions depends only on angle)\cite{3}. In contrast to the detection of high frequency waves, these detection occurs via the accumulation of signals over many years. As mentioned before, these detection are useful to know about galaxies mergers and black hole dynamics and to explore fundamental physics like measuring the cosmological constant of the Universe\cite{15} and testing general theory of relativity.

\begin{acknowledgements}
I would like to thank Prof. Gavin Brennen for his useful comments. Similarly, I would like to thank Elija Timalsina for her feedbacks after reading this article carefully.
\end{acknowledgements}

\end{document}